\RequirePackage{fix-cm}
\documentclass[smallextended]{svjour3}       
\smartqed  
\usepackage{graphicx}
\usepackage{natbib}
\begin{document}
\title{Detection of Crab radiation with a meteorological balloon borne phoswich
detector 
\thanks{This work is partially supported by the Science and Engineering
Research Board, India grant number EMR/2016/003870
}
}


\titlerunning{Phoswich detector on board meteorological balloon}        

\author{Ritabrata Sarkar \and
        Sandip K. Chakrabarti \and
        Debashis Bhowmick \and
        Arnab Bhattacharya \and
        Abhijit Roy
}

\authorrunning{Sarkar et~al.} 

\institute{Ritabrata Sarkar 
           \and Sandip K. Chakrabarti 
           \and Debashis Bhowmick 
           \and Arnab Bhattacharya 
           \and Abhijit Roy 
           \at Indian Centre for Space Physics, 43 Chalantika, 
           Garia Station Rd., Kolkata 700084\\
           Tel.: +91-33-24366003\\
           Fax: +91-33-24622153\\
           \email{ritabrata.s@gmail.com}
}


\maketitle

\begin{abstract}
We use existing light weight balloon facility of Indian Centre for Space Physics
to detect the X-ray radiation from Crab pulsar with a phoswich detector. We
present the design considerations and characterization of the detector used for
this purpose. We model the background radiation in the detector environment at
various altitudes and use this in spectral analysis. The background radiation
level and limitations on the detector allowed us to calculate minimum detection
limit for extrasolar radiation sources with our set up.



\end{abstract}

\section{Introduction}
\label{sec:intro}
Indian Centre for Space Physics (ICSP) has been pursuing an independent space
program with light-weight meteorological balloons for more than a decade
\citep[collectively denoted as \textit{C1117}]{chak11, chak13, chak14, chak15,
chak17}. In this new paradigm of the space exploration with light weight
detectors clearly have merits and difficulties. However, despite their
limitations a great deal of science can be done. The payloads used in these
`Dignity' series missions are about 5 kg weight containing mainly charge
particle or X-ray detectors along with some auxiliary systems. Details of such
experimental initiative can be found in \citet{chak17, sark18} and references
therein.

Apart from various constraints, two main challenges to face in this type of
experiments are: excess background radiation in the atmosphere and availability
of limited time to observe a specific source. The minimum detectability or
sensitivity of the detector is limited by both these effects and one has to
carefully choose the experimental parameters and conditions to get the highest
possible exposure of the source and lowest possible the background. One of the
adverse characteristics of using light weight payloads on board single
meteorological balloons is that the in the absence of ballast and valve systems,
the balloon goes up to a certain height, gets ruptured at the burst height and
then steadily comes down. Since there is not really a cruising level of the
payload, it is always under diverse background environment and at the same
time radiation from the (external) source also undergoes different amount of
absorption due to the ever changing residual atmosphere.

Moreover, due to the weight constraint on the payload, we cannot deploy a
pointing device in the payload to lock the source in the detector direction.
Thus the payload is free to rotate about its axis which is along the string
attaching the payload to the balloon. The payload is also under local turbulence
due to wind pattern and its interaction with the balloon itself. One needs to
consider these effects along with the payload altitude during the analysis of
the radiation data from the source. These considerations constrain the design of
the collimator of the detector to provide an optimum Field of View (FoV) to the
detector so that a balance between background counts and exposure time may take
place.

The nature of the background noise depends on the payload altitude, observation
time and other space weather effects like solar activity etc. and discussed
in more details in \citet{bazi98, sark17} and \textit{C1117}. However, along
with the external background we also have to take detector internal background
into account \citep{pete75}. The estimation and minimization of the background
is particularly important for the observation of astronomical objects which are
very faint and the signal-to-noise ratio could be quite poor. In the present
work, we demonstrate how modeling of the background radiation could be
done using a strong and standard astronomical X-ray source, namely, the 
Crab pulsar. We use a 5'' phoswich scintillator detector for this purpose.
Since the discovery of Crab in 1968 by radio telescope \citep{stae68} and
subsequently in gamma-ray band \citep{vass70} through balloon-borne experiment,
many major astronomical experiments continue to observe the source. Being one of
the most powerful extrasolar sources emitting quite stable flux density with
stable spectral shape, this source has acted as a standard calibrator.
Nevertheless, the source remained a fascinating object to the astronomers and
the complete physical processes responsible for its emission still remain to be
understood \citep{kirs05, vada17}. In the same spirit, we also choose this
source as the candidate for observation to guide us the way the backgrounds
should be eliminated and to validate our new paradigm for X-ray astronomy.

The developmental works of this paradigm are already described in \textit{C1117}
and we do not repeat them here. With X-ray detectors, the Dignity series so far
has flown 109 missions many of which were to test various modules. In
\S\ref{sec:instr}, we briefly describe a general mission present the detector
design. We discuss detector characteristics such as calibration and resolution
and its response functions in \S\ref{sec:char} and the effects of the residual
atmosphere in \S\ref{sec:atms}. In \S\ref{sec:bkg}, we discuss the detector
background characteristics at different altitudes and formulate an empirical
model to calculate the background and determine the sensitivity of the detector
at various altitudes. In \S\ref{sec:crab} we present the results of radiation
detection from the Crab pulsar and finally we conclude our work in
\S\ref{sec:conc}.

\section{Experiment and instrumentation}
\label{sec:instr}
A scintillator detector in phoswich combination was flown on board balloon in
Dignity mission. The phoswich detector consists a NaI(Tl) crystal of 3 mm
thickness and 116 mm diameter. Beneath the NaI crystal there is a 25 cm thick
CsI(Na) crystal of same diameter to provide the active background noise
cancellation using anticoincidence techniques. The scintillation light produced
in the crystals are read out by the Photo-Multiplier Tube (PMT) and their
origins are distinguished from the corresponding pulse shape. Similar detectors
were used in the RT-2 experiment \citep{rao11} on board CORONAS-PHOTON satellite
to study solar activity and other galactic and extra-galactic sources. More
details of the detector are in \citet{debn11}.

\begin{figure}[h]
  \centering
  \includegraphics[width=0.4\textwidth]{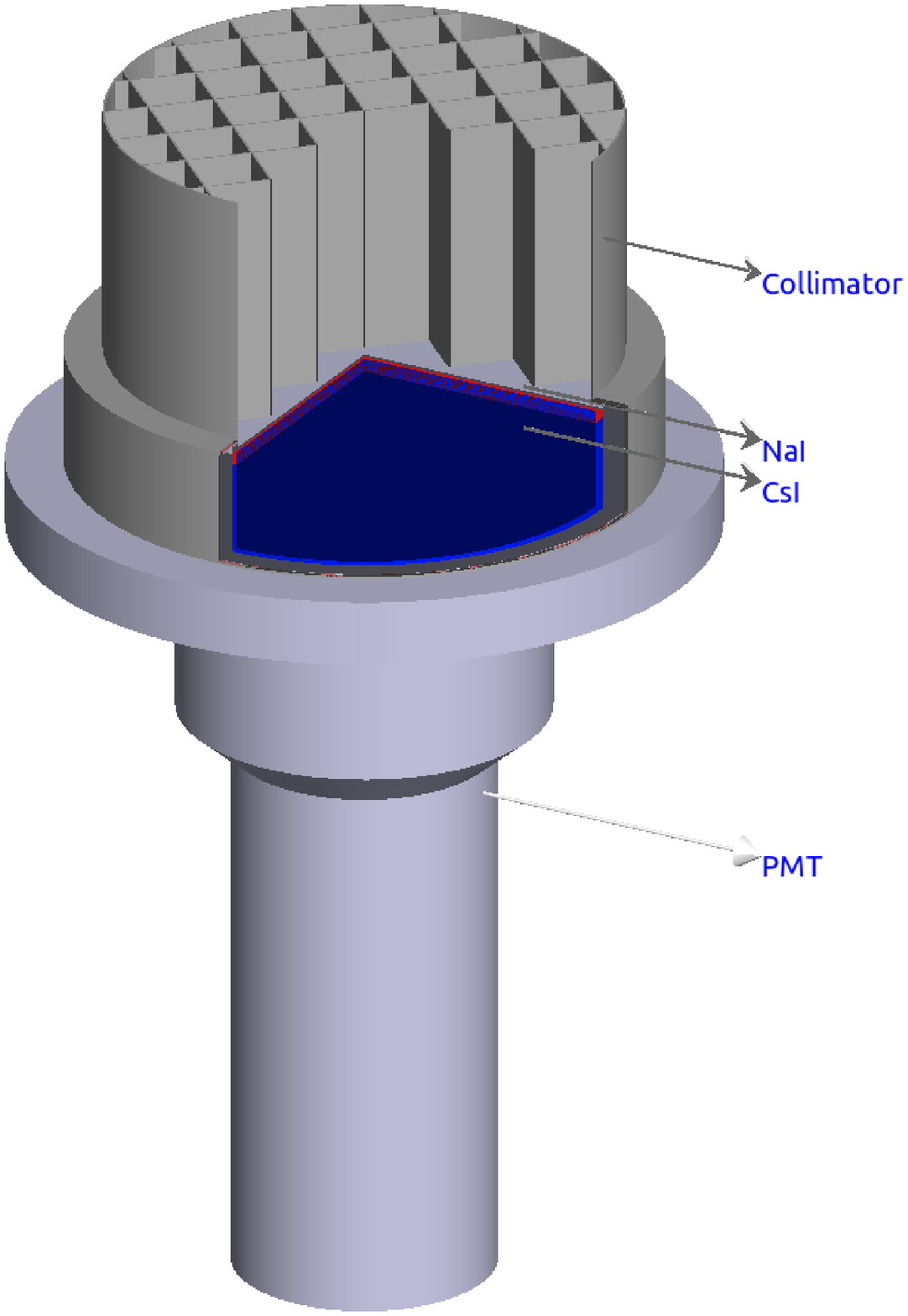}
  \includegraphics[width=0.59\textwidth]{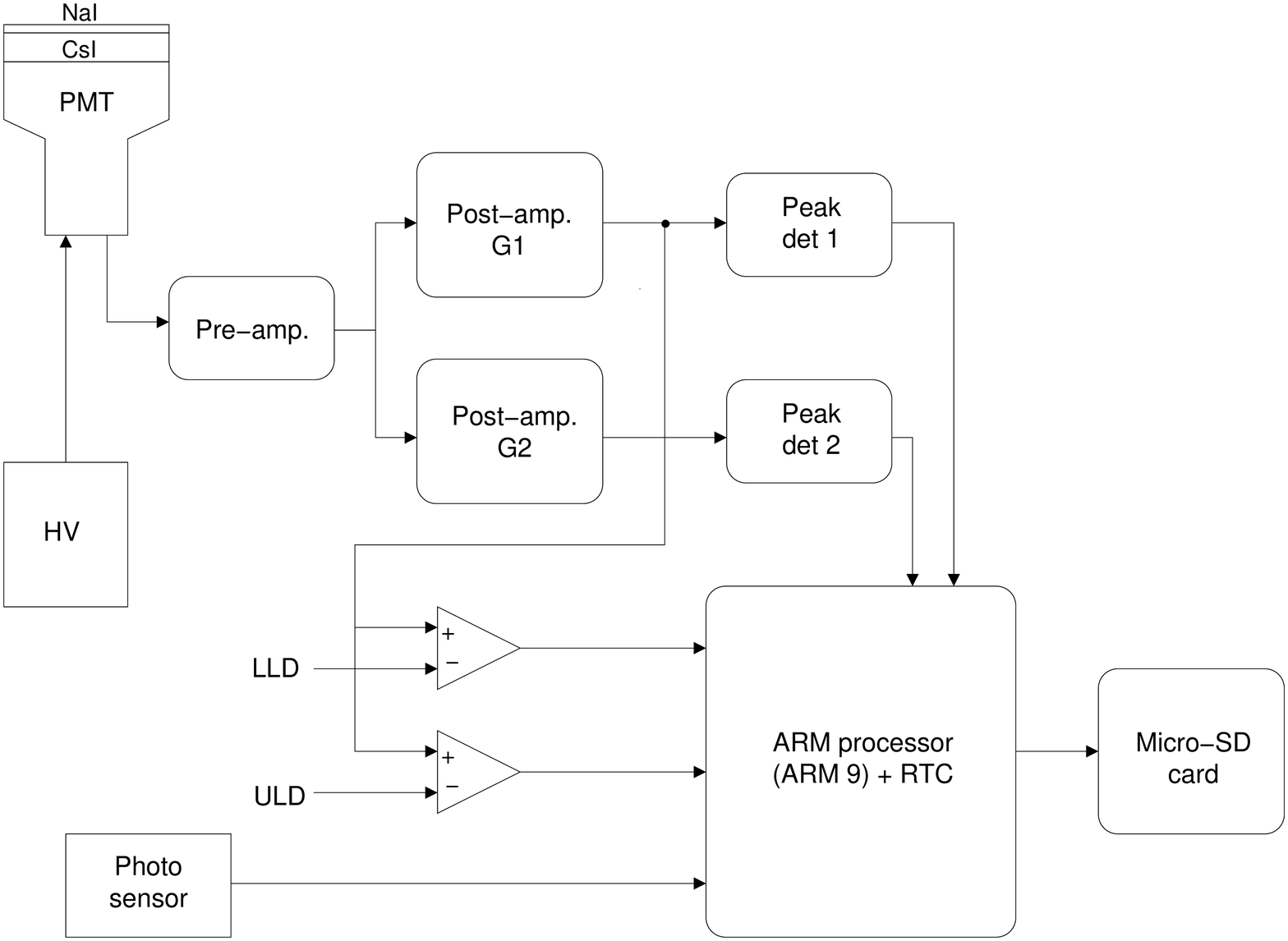}
  \caption{(Left:) a schematic diagram of the phoswich detector along with the
  PMT, collimator and shielding. (Right:) block diagram representation of the
  electronics and readout system of the detector.}
   \label{fig:det}
\end{figure}

The payload with the same detector was redesigned with optimized shielding,
collimator and readout system to suit our experimental environment and science
goals. We shielded the crystal part with 1 mm tin and 0.2 mm copper underneath
to prevent the low energy background where the detector is most sensitive. We
put no extra shielding at the lower part containing the PMT due to weight
constraints, as the low energy photons from this direction into the detector
will first interact in CsI and will be discarded by the anticoincidence logic.
The collimator is made of 0.5 mm lead to provide $\sim$ 15$^{\circ}$ FWHM FoV.
This is the optimum FoV calculated considering the lack of pointing device which
could lock the source in the FoV for longer exposure. But since our payload is
free to rotate around its axis and susceptible to external jittering, the
exposure depends on these rotation and jittering effects and also on the motion
of the source (due to earth's rotation) through the detector FoV. So we need to
keep the FoV wider to have longer exposure but in turn it reduces the
sensitivity of the detector as it allows larger external background. A schematic
diagram of the detector along with the collimator and shielding is shown in Fig.
\ref{fig:det} (left panel) and the block diagram of the readout scheme is given
at the right panel of the same figure.

The detector and the electronic readout system are powered from battery provided
in the payload. The signal amplification system is achieved by two post
amplifiers. One (G1) for the low energy part giving $\sim$ 15-100 keV energy
range for most of the radiation sources of our interest which shows
significant activity in this energy range. The other post amplifier (G2) can
provide energy detection up to $\sim$ 2000 keV to detect other high energy
phenomena like TGF etc. In the scope of this current work we will only consider
the data from G1. More details of the detector configuration, operation and
readout system is described in \citet{bhow18}.

The principal purpose of this experiment is to measure the radiation from the
standard astronomical source: Crab pulsar. So we scheduled the date and time of
the experiment to have the source inside the FoV of the detector when it is near the
top of the atmosphere and without much tilting the viewing axis of the detector
from zenith to discard the reduction of effective exposure time due to the
payload rotation. The summary of the mission related information are given in
Table \ref{tab:mis} and more detailed methodology of the experiment and the
payload can be found in \citet{chak17}.

\begin{table}[h]
 \begin{center}
   \begin{tabular}{rp{4cm}p{8cm}}
    \hline
    \vspace{1mm}
    Main detector on board & 116 mm diameter phoswich detector with 3 mm NaI
    and 25 mm CsI crystal.\\
    \vspace{1mm}
    Collimator & made of 0.5 mm thick lead (15$^{\circ}$ FWHM FoV).\\
    \vspace{1mm}
    Other ancillary instruments & GPS system, payload tracker system, attitude
    measurement system.\\
    \vspace{1mm}
    Total payload weight & 5.7 kg.\\
    \vspace{1mm}
    Payload carrier & one meteorological plastic balloon (length: $\sim$ 25 m,
    weight: 7.55 kg).\\
    \vspace{1mm}
    Date of launch & 12 May, 2016.\\
    \vspace{1mm}
    Time duration & 6:12 -- 9:13 UT.\\
    \vspace{1mm}
    Place of launch & Muluk, W.B., India (Lat: 23.6468, Lon: 87.7151).\\
    \vspace{1mm}
    Maximum height attained & 41.05 km (at 7:23 UT).\\
    \hline
   \end{tabular}
   \caption{Mission particulars of the experiment.}
   \label{tab:mis}
 \end{center}
\end{table}

\begin{figure}[h]
  \centering
  \includegraphics[width=0.6\textwidth]{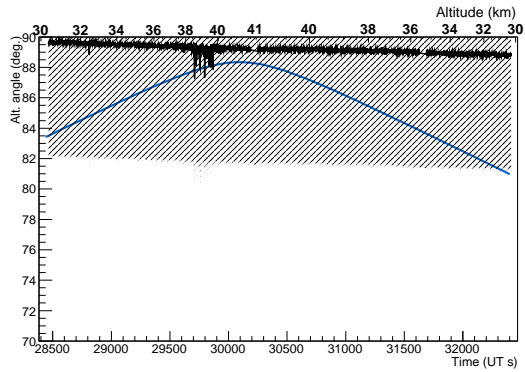}
  \caption{Direction of the detector along with sky coverage by the FoV (black)
  and position of Crab (blue) w.r.t time and altitude above 30 km. The FoV of
  the detector is shown by the shaded area.}
   \label{fig:pos}
\end{figure}

We show the position of Crab location in altitude angle in the local sky with
time and payload altitude in Fig. \ref{fig:pos} (only above 30 km). We also plot
the detector direction and the sky coverage of the detector FoV at each instant.
It is evident from the plot that Crab is inside the FoV for the entire time when
the payload is above 30 km and we also calculate the effective area provided by
the detector for the position of the source at each instant which is necessary
for the spectral analysis of the radiation from the source.

\section{Detector characteristics}
\label{sec:char}
We studied and optimized the detector characteristics for these types of missions
through several tests in the laboratory and Monte-Carlo simulations. The details
of the laboratory tests done on the detector to study the effect of diverse
atmospheric conditions such as temperature, pressure etc. are described in
\citet{chak17, bhow18} and \textit{C1117}.

\subsection{Calibration and resolution}
\label{ssec:cal}
For calibration of the detector we use laboratory radiation sources: Eu152,
Ba133 and Am241 which have several lines inside the sensitive energy range of
the detector to give a linear channel-energy relationship. We also calculate the
detector resolution at different energies which we find to vary as
$6.62\,E^{-0.85}$. We use this relation for binning of the detector energy
range for spectral studies.

\subsection{Response Matrix Function}
\label{ssec:rmf}
\begin{figure}[h]
  \centering
  \includegraphics[width=0.6\textwidth]{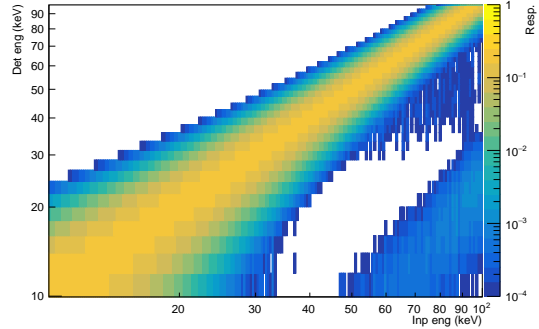}
  \caption{RMF of the detector for energy depositions in NaI crystal calculated
  from simulation and laboratory test.}
   \label{fig:rmf}
\end{figure}

The overall response of the detector to the external radiation is divided in
two components: the Response Matrix Function (RMF) and Ancillary Response Function
(ARF). RMF contains the differential probability of a photon incident
on the crystal to be detected in certain energy bin. To generate the RMF of the
detector (and for other simulations used in this work) we considered a
Monte Carlo simulation of the detector using Geant4 simulation toolkit (version
10.3.1) \citep{agos03} and the energy resolution of the detector at different
energies obtained from the laboratory test. For the simulation we considered a
flat distribution of 10$^6$ photons in the energy range of 10-100 keV incident
on the NaI crystal from above the detector and calculated exclusive energy
deposition in NaI crystal only. We modulated the deposited energy in
the crystal according to the corresponding resolution of the detector and
populated the response matrix. For the electromagnetic interactions of photons
in the crystal we considered ``em-standard'' physics list defined in Geant4. The
RMF for the NaI crystal of the detector is shown in Fig. \ref{fig:rmf}.

\subsection{Ancillary Response Function}
\label{ssec:arf}
\begin{figure}[h]
  \centering
  \includegraphics[width=0.45\textwidth]{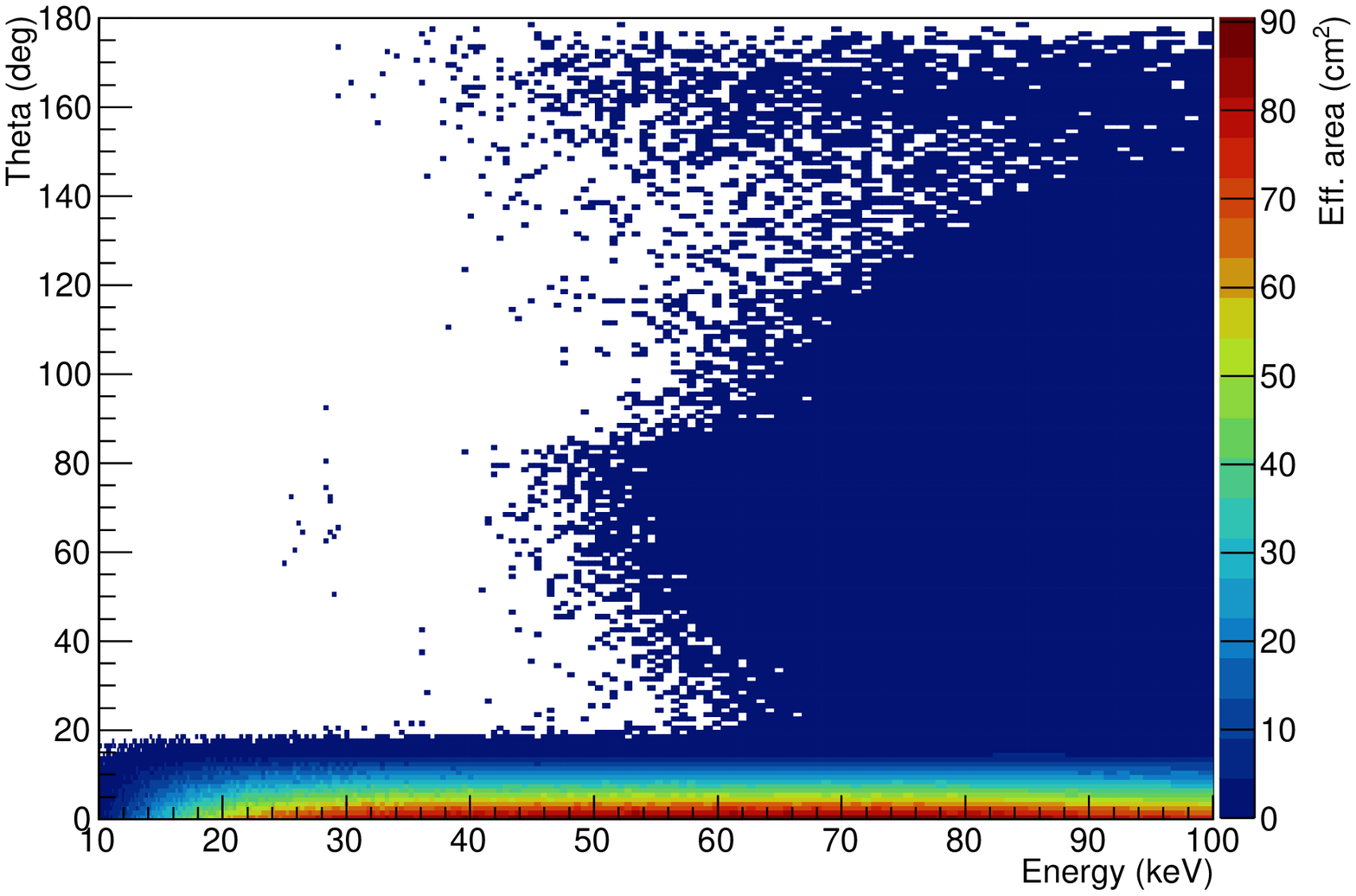}
  \includegraphics[width=0.45\textwidth]{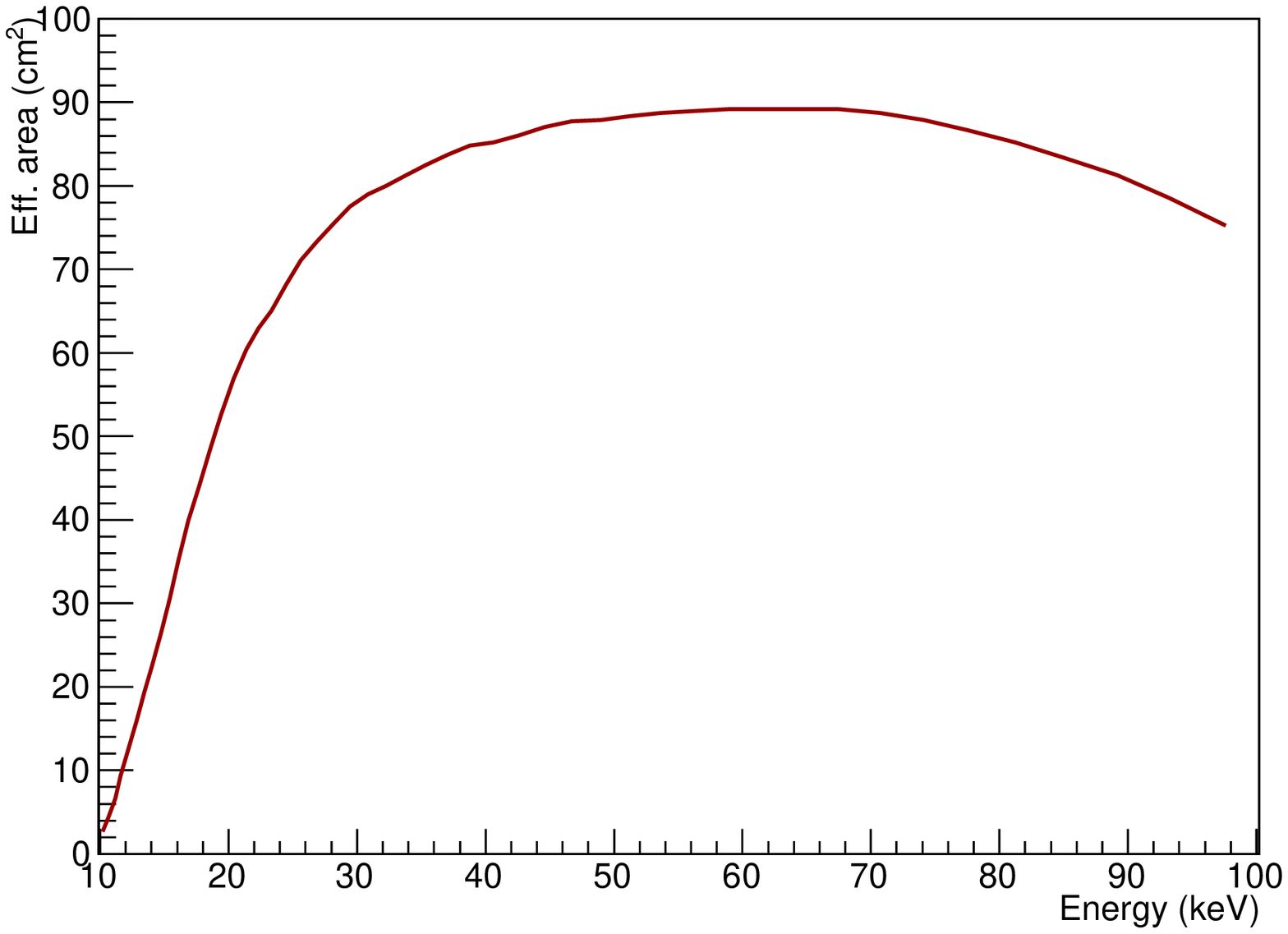}
  \caption{(Left:) Effective area distribution of the detector at different
  $\theta$ w.r.t energy. (Right:) Effective area plot at $\theta = 0^{\circ}$.}
   \label{fig:arf}
\end{figure}

The ARF consists of the effective area of the detector for different energies and
incident angles. We simulated the detector configuration with proper shielding
and collimator with incident photons in the energy range of 10-100 keV from
4$\pi$ solid angle to the detector with random directional distribution. The
number of photons entering into the detector will depend on direction of the
photons due to position of the crystals and shielding and collimator
configuration. The area subtended by the detector crystal also change with the
incident angle. Considering these, we calculate the effective area (which
also includes the efficiency) of the detector and generate a 3 dimensional
matrix with these information for different energies, altitude angle ($\theta$)
and azimuth angle ($\phi$). The ARF information for various energies at
different $theta$ is shown in Fig. \ref{fig:arf} in the left panel while the same
information for $\theta = 0^\circ$ is plotted at the right panel. 

\section{Atmospheric response}
\label{sec:atms}
There remains a residual atmosphere at the balloon height causing the
absorption and modification of radiation spectrum of the extraterrestrial
sources. These must be taken into account to obtain the actual source spectrum.
Moreover from Fig. \ref{fig:pos} we see that the detector undergoes
different amounts of residual atmosphere due to time dependent payload altitude.
So, for the spectral analysis with longer exposure time we must dynamically
correct for this atmospheric absorption effect. For this purpose we generated
a response matrix function and absorption function for the residual atmosphere
using Geant4 simulation.

\begin{figure}[h]
  \centering
  \includegraphics[width=0.45\textwidth]{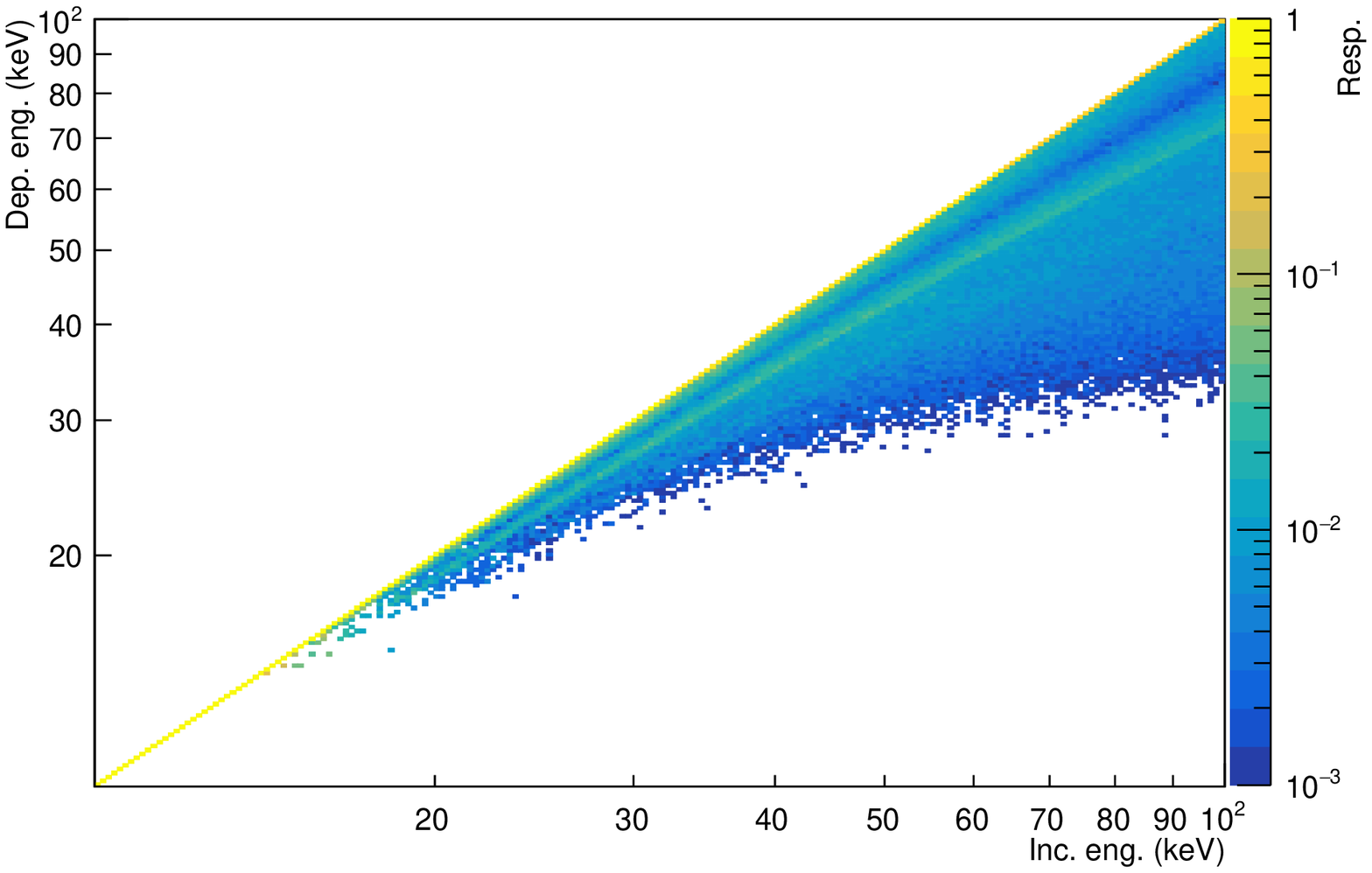}
  \includegraphics[width=0.45\textwidth]{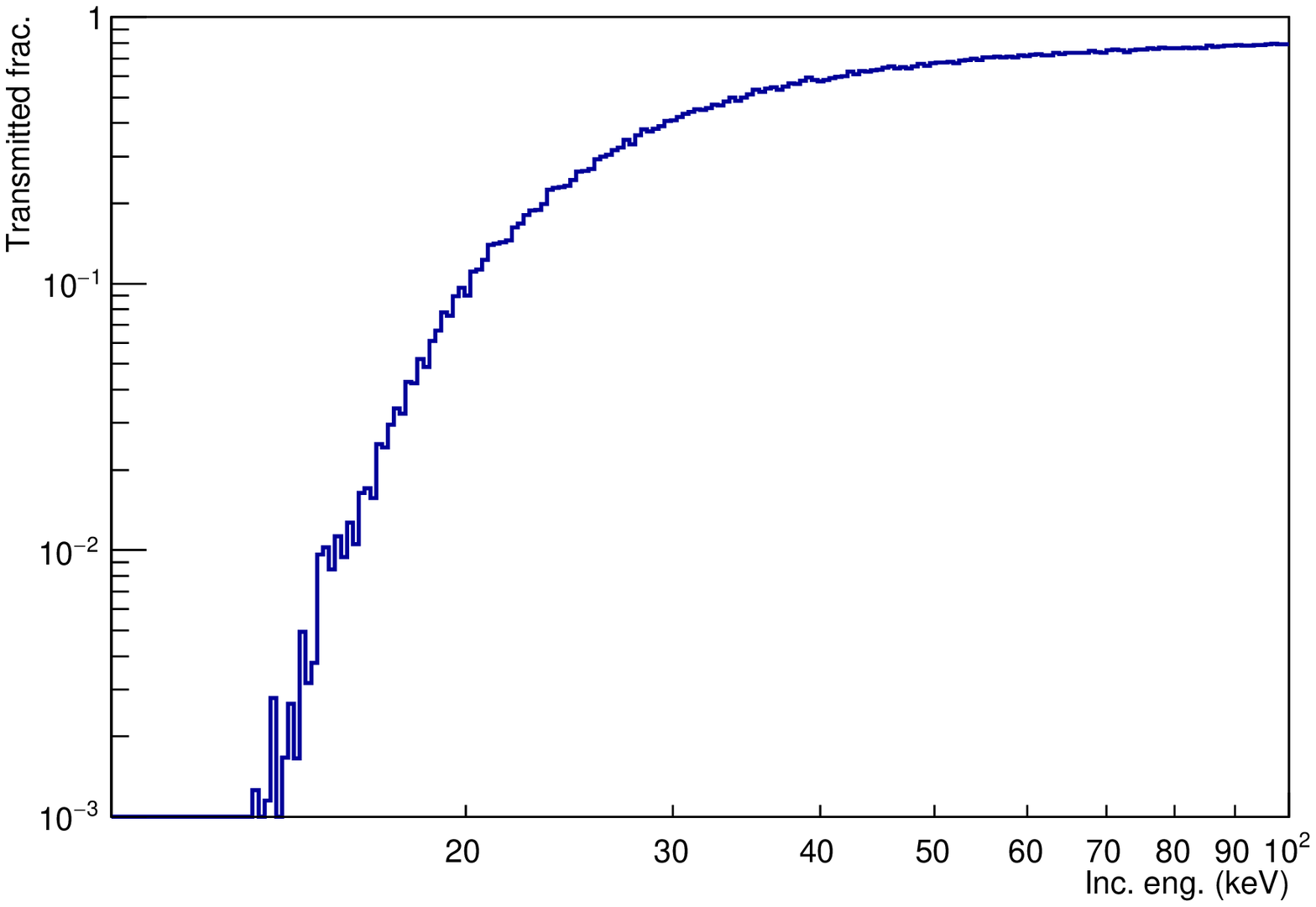}
  \caption{(Left:) Atmospheric response matrix for the residual atmosphere at
  an altitude of 38 km. (Right:) Transmission fraction for photons of
  different energies at the same altitude.}
   \label{fig:atmrsp}
\end{figure}

To define the atmosphere we considered the NRLMSISE-00 standard atmospheric
model parameters \citep{pico03} with parallel layers of 1 km thickness up to
50 km. We used incident photons in the energy range 10-100 keV perpendicular to
the layers from the top of the atmosphere and calculated the response function
and the absorption function at various altitudes up to 42 km. An example of
the response and absorption functions at an altitude of 38 km is shown in Fig.
\ref{fig:atmrsp} for normal incidence of radiation. However depending on the
source position, the incident angle ($\theta_i$) may vary and the absorption will
be affected by this angle and for larger angles the flat atmospheric layer
approximation will fail. But for small incident angles (like $\le 15^{\circ}$ in
this case which is limited by the FoV of the detector) the absorption may be
approximated to vary as $1/\cos{\theta_i}$. We need to use these atmospheric
response and absorption functions in addition to the detector response for the
analysis of external radiation.

\section{Background calculation}
\label{sec:bkg}
The sensitivity of the detector in these types of experiments is highly affected by
the atmospheric background due to Cosmic Ray (CR) interactions apart from the
primary cosmic background and those which are intrinsic to the detector. Extensive
calculations of background noise for a similar detector was done by
\citet{sark11} for experiments in satellite environment. In the present experiment,  
the situation is different due to the presence of atmosphere which makes the
computation more complicated. But the fact of not having much high-Z material in the
vicinity of the detector means less secondary background from the payload is produced.

There is also a specific problem for these types of experiments regarding the
background estimation which one has to overcome. We cannot have actual
measurement of simultaneous background when the source is inside the detector
for a long time. In this case we must rely on the accurate modelling of the
background in that condition. This background has an external part which
contains cosmic rays and secondary radiations due to their interaction with the
atmosphere. The other part of the background, intrinsic to the detector may
again depend on the external background flux and is caused by the activation
and spallation of the detector materials. The primary and secondary CR
background depends on the geomagnetic latitude and altitude of the experiment.
This CR background on the other hand modulated by space environment condition
mainly due to solar activity and is discussed in \citep{sark17} for similar
latitude and energy range.

\subsection{Trigger efficiency for background components}
\label{ssec:trg}
The external background components at the balloon flight altitude consist of
charged particles (protons, alpha, electrons, positrons) and gamma-rays
in the cosmic-ray flux and muons from the atmospheric interactions of the primary
CRs. \citet{mizu04} provides an extensive discussions on the energy dependent
flux of these components and their atmospheric depth and angular dependency at
the balloon height. 

We simulated our detector to calculate the trigger efficiency for each of the
external background components distributed in 4$\pi$ solid angle of the
detector. The trigger efficiency of the detector (for incident energy range 10
keV - 100 MeV) and the integrated background flux in the energy range of
10-100 keV due to various background components at a height of 38 km is given
in Table \ref{tab:bkg}. From the calculation we notice that the most effective
component of the background is due to gamma-ray for this type of
payloads. The other particles produce negligible trigger in the detector due
to thin primary crystal and rejection due to anti-coincidence technique and
therefore we do not take them into account. Moreover, the primary gamma-ray flux
in CR is less than one order in magnitude to that of the downward secondary
photons produced in atmosphere \citep{mizu04}, so we consider only secondary
gamma-ray in the atmosphere as the principal background source.

\begin{table}[h]
 \begin{center}
   \begin{tabular}{lcc}
    \hline
    \vspace{1mm}
    Bkg. comp. & Trg. eff. & Int. flux\\
    \hline
    \vspace{1mm}
    Gamma & $0.174$ & $0.54$\\
    \vspace{1mm}
    Proton & $1.8\times10^{-5}$ & $7\times10^{-6}$\\
    \vspace{1mm}
    e$^-$ & $6.9\times10^{-4}$ & $3\times10^{-3}$\\
    \vspace{1mm}
    e$^+$ & $6.6\times10^{-4}$ & $4\times10^{-3}$\\
    \vspace{1mm}
    $\mu^-$ & $1.9\times10^{-3}$ & $4\times10^{-5}$\\
    \vspace{1mm}
    $\mu^+$ & $2.9\times10^{-4}$ & $1\times10^{-5}$\\
    \hline
   \end{tabular}
   \caption{Trigger efficiency (for incident particles in energy range in 10
   keV - 100 MeV) and integral flux (counts/cm$^2$/s) in the detector in the energy
   range of 10-100 keV for various cosmic-ray background components.}
   \label{tab:bkg}
 \end{center}
\end{table}

\subsection{Detector background modelling}
\label{ssec:bkgmod}
From Fig. \ref{fig:pos} we see that above 30 km the Crab was inside the FoV of
the detector. So there is no other way to directly measure the simultaneous
background and we need to rely on the calculated background of the detector.
For this reason we developed an empirical model of the detector background. For
residual atmosphere less than 100 g~cm$^{-2}$ the downward secondary gamma-ray
flux is proportional to atmospheric depth and upward flux is almost constant
\citep{thom74}. So we considered to fit the detector backgrounds from 20 km
(55.8 g~cm$^{-2}$) till 30 km in each layer of 1 km with the empirical model and
extrapolate the model parameters for higher altitudes to calculate the
corresponding background. 

For external background due to downward and upward secondary gamma-rays we
consider the flux distribution according to the power law \citep{mizu04}:
\begin{equation}
A\,\left(\frac{E}{keV}\right)^{-1.34} counts\,s^{-1}\,m^{-2}\,sr^{-1}\,keV^{-1}
\label{eqn:bkg} 
\end{equation}
where the flux normalization $A$ varies with the atmospheric depth and depends
on the geomagnetic latitude of the experiment and solar activity. The zenith
angle dependence of the flux was taken into account as described in
\citet{mizu04}.

There are several background sources  which are fundamentally internal to the detectors
due to radio activity stimulated by CRs \citep{grub96}. Lines coming from K
X-ray at 25-30 keV from spallation of tellurium, 58 keV inelastic scattering
gamma in iodine are clearly present in the background spectrum.

\begin{figure}[h]
  \centering
  \includegraphics[width=0.6\textwidth]{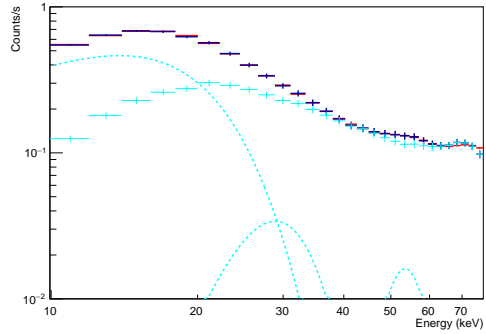}	
  \caption{Detector background at 30 km (red) and the calculated background
  according to the empirical model (blue). The model components are shown (in
  cyan) which consist of three Gaussian (dashed line) for intrinsic background
  and a power law component for external secondary gamma-ray modified by the
  detector response (see text for details).}
   \label{fig:bkg}
\end{figure}

We fitted the detector background using Equation \ref{eqn:bkg} convoluted by the
detector response for different incident angles with one Gaussian to account
for the low energy noise near $\sim$ 12 keV and two additional Gaussian lines
around 30 and 60 keV. We extracted the model parameters for different
atmospheric depths and fitted their variations. We found the flux normalization
$A$ to vary exponentially with atmospheric depth while other parameters vary
linearly. In Fig. \ref{fig:bkg} we show the comparison of the detector
background and the modelled background at 30 km height. Using these parameters
we are able to calculate the background at higher altitudes when Crab is inside
the detector FoV.

\subsection{Sensitivity of the detector}
\label{ssec:sens}
\begin{figure}[h]
  \centering
  \includegraphics[width=0.6\textwidth]{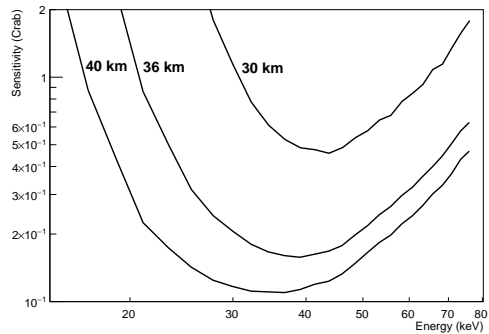}	
  \caption{Detector sensitivity with energy in units of Crab radiation flux
  calculated at three different heights of 30, 36 and 40 km.}
   \label{fig:sen}
\end{figure}

From the background flux in the detector we calculate the minimum detectable
limit or sensitivity of the detector at various altitudes in the atmosphere. In
Fig. \ref{fig:sen} we have shown, for example, the sensitivity of the detector
at three different altitudes of 30, 36 and 40 km. This calculation is done
considering 5 minutes of detector exposure time and signal-to-noise ratio of
3 which corresponds to about 99.8\% confidence level. The sensitivity is expressed
in units of Crab radiation flux modified by the residual atmosphere at that
altitude. The calculation affirms that detection of extrasolar sources with
Crab like spectrum and a few hundred mCrab flux are possible in a limited hard
X-ray energy range.

\section{Crab data analysis}
\label{sec:crab}
\begin{figure}[h]
  \centering
  \includegraphics[width=0.6\textwidth]{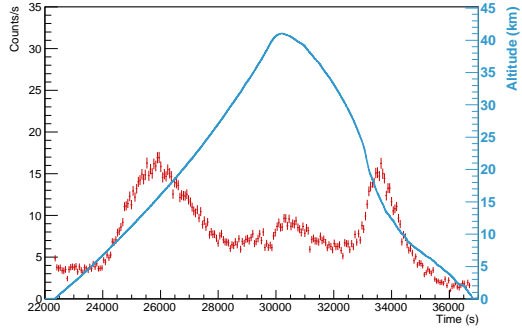}	
  \caption{25-60 keV light curve of the overall mission data showing the
  radiation from Crab above 30 km. Payload altitude is plotted simultaneously
  to show the detector position.}
   \label{fig:lc}
\end{figure}

The radiation flux detected in the entire mission flight in the energy range of
25-60 keV is shown in Fig. \ref{fig:lc}. In the light curve the first and last
peaks are from atmospheric radiation at Regener-Pfotzer maximum due to secondary
cosmic rays and has been discussed in \citet{sark17}. The peak at the middle is due to
the radiation from Crab pulsar and is evident from Fig. \ref{fig:pos}. There is
a little dip in the light curve at around 29 ks which is due to turbulence in
the carrier during the balloon rupture and has been confirmed from the camera
installed on the payload and from its attitude measurement.

For the spectral analysis of the radiation from the Crab we considered a total
of 1178s exposure data which spanned in the altitude range of 38-41 km. For
the background data we used the calculated data described in
\S\ref{ssec:bkgmod}. We considered a weighted average of the background in each
layer (of 1 km) in the altitude range according to the time taken by the
payload to cross each layer.

The Crab spectrum is fitted by the absorbed power-law model \citep{kirs05}:
\begin{equation}
  S(E_i) = A\,E_i^{B}\,exp(-0.38 \sigma(E_i))
  \label{eqn:crab}
\end{equation}
where, $S(E_i)$ is in units of photons/cm$^2$/s/keV and $\sigma(E_i)$ is H
column cross-section in units of cm$^2$. This source spectrum is modified by the
atmospheric absorption function ($A_{abs}(E_i)$), atmospheric response function
($A_{rsp}(E_i, E_a)$), detector ARF ($D_{arf}(E_a)$) and detector RMF
($A_{rsp}(E_a, E_d)$) to fit the detected source spectrum $F(E_d)$.  
\begin{equation}
  F(E_d) = A_{rsp}(E_a, E_d) . \, D_{arf}(E_a) . \, A_{rsp}(E_i, E_a) . \,
  A_{abs}(E_i) . \, S(E_i)
  \label{eqn:mod}
\end{equation}

\begin{figure}[h]
  \centering
  \includegraphics[width=0.6\textwidth]{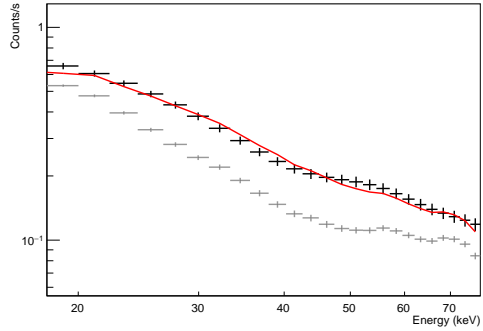}	
  \caption{The spectrum of Crab radiation (source + background) as seen by the
  detector (black) along with the calculated detector background (gray). The
  spectrum is fitted by using an absorbed power law modified by the atmospheric
  and detector responses.}
   \label{fig:spec}
\end{figure}

The best fit of the spectrum in the energy range of 20-75 keV gives $A$ = 12.36
$\pm$ 4.66 photons/cm$^2$/s/keV and $B$ = -2.41 $\pm$ 0.10 with $\chi^2$ = 26.12
for 23 NDF. The detected spectrum of Crab radiation along with the calculated
background and the model fit is plotted in Fig. \ref{fig:spec}.

\section{Conclusions}
\label{sec:conc}
Indian Centre for Space Physics, through its Dignity missions, has been
exploring near space using meteorological balloons for a decade (e.g.,
\citet{chak14, chak15, chak17, sark17}) and preliminary studies suggest that a
great deal of valuable scientific experiments could be carried out with these
low-cost Missions. However, unlike experiments with larger balloons and massive
payloads, where pointing to a target is possible to raise exposure time and the
signal to noise ratio, in our case, pointing is not done explicitly and each
photon is tagged with an attitude from which source information is obtained
through post-processing. Otherwise corrections due to atmospheric effects and
background radiation remain a concern for both the cases.

In the present paper, we went into details of how to eliminate effects of
background radiation while studying cosmic X-ray sources. For a case study, we
chose a strong and steady X-ray source, namely, the Crab pulsar and show that by
systematically eliminating the background noise we recover the spectrum very
well. We computed the limit on the sensitivity of the detector. We presented the
techniques to estimate the detector characteristics and also estimated the
environmental effects on the experiment, such as, the effects of the residual
atmosphere. Presently, we find that sources with brightness at ten per cent
level as that of the Crab may also be detected by these low-cost missions.
However, the sensitivity of could be increased further by using lighter
detectors with larger area and using better shielding and collimation.

\begin{acknowledgements}
The authors would like to thank the ICSP balloon team members, namely,
Mr. S. Midya, Mr. H. Roy, Mr. R. C. Das and Mr. U. Sardar for their 
dedications during the crucial mission operations such as launching and recovery.
This work been done under partial financial support from the Science and
Engineering Research Board (Science and Engineering Research Board, Department
of Science and Technology, Government of India) project no. EMR/2016/003870. We
also thank Ministry of Earth Sciences (Government of India) for partial
financial support.
\end{acknowledgements}


\end{document}